# Keeping it Authentic: The Social Footprint of the Trolls' Network

Ori Swed[1], Sachith Dassanayaka[2], & Dimitri Volchenkov[3]

[1]Department of Sociology, Anthropology, and Social Work, Texas Tech University

[2]Department of Mathematics & Computer Science, Wittenberg University

[3]Department of Mathematics & Statistics, Texas Tech University

**Abstract**—In 2016, a network of social media accounts animated by Russian operatives attempted to divert political discourse within the American public around the presidential elections. This was a coordinated effort, part of a Russian-led complex information operation. Utilizing the anonymity and outreach of social media platforms Russian operatives created an online astroturf that is in direct contact with regular Americans, promoting Russian agenda and goals. The elusiveness of this type of adversarial approach rendered security agencies helpless, stressing the unique challenges this type of intervention presents. Building on existing scholarship on the functions within influence networks on social media, we suggest a new approach to map those types of operations. We argue that pretending to be legitimate social actors obliges the network to adhere to social expectations, leaving a social footprint. To test the robustness of this social footprint we train artificial intelligence to identify it and create a predictive model. We use Twitter data identified as part of the Russian influence network for training the artificial intelligence and to test the prediction. Our model attains 88% prediction accuracy for the test set. Testing our prediction on two additional models results in 90.7% and 90.5% accuracy, validating our model. The predictive and validation results suggest that building a machine learning model around social functions within the Russian influence network can be used to map its actors and functions.

The 2016 US presidential elections introduced a quarrelsome political discourse and a dramatic outcome. They were also accompanied with a massive Russian interference campaign that aimed at undermining democratic processes [1]. A campaign that aggressively promoted Russian agenda through the promotion of pro-Russian candidates, and attempts to influence voters and public [2]. This intervention was an information operation targeting potential voters on social media to influence their political worldview and, consequently, their political decision-making and participation in the democratic process [3]. The operation included the creation of a wide array of complementing fake accounts and personas on social media platforms [4, 1]. Those accounts and personas, animated by Russian operatives, impersonate regular American folks from rural Iowa to the suburbs of Atlanta [5]. Building on the anonymity social media platforms offer, Russian operatives could conceal their real identity and agenda to gain the trust of their unsuspected American audiences. The elusiveness of this tactic caught the country, and security agencies in particular, by surprise. How do you identify a troll? How do you isolate a malicious account from the rest? In an environment that often cultivates polemic and hyperbolic behavior, how can we separate an angry conservative or liberal from an inflammatory foreign provocateur? Those challenges invited experts and scholars to try to map this operation, the trolls' network, and the tactics used [5, 6, 7, 8, 9].

A growing body of research indicates that some of the functions in the Russian trolls' network are structural and repetitive, and as such, allow us to identify and map actors in the network. We argue that some of those patterns have a social dimension, not merely technical. As an artificial network that eyes a collective goal, the Russian network bears similarities to an astroturf, a fake social movement that promotes paid agenda in concert [10, 11]. The troll network demonstrates similar traits as it attempts to push its chosen agenda in a coordinated yet unnoticeable effort [12]. It tries to present an authentic and natural façade to convince the audience they converse with a real native, that her/his opinion is valid and should be valued, and not a fake persona or a bot. To do so, the fake persona is required to maintain a coherent and convincing social appearance, one that differentiates an individual from an organization and that will fit with the presentation of self. This social footprint confines the accounts' flexibility and generates a pattern of behavior. Namely, a news account reporting on Lubbock, Texas cannot change abruptly to an account of a suburban housewife from Idaho that discusses gardening; it will diminish its credibility. We essentially argue that the "social footprint" is a prerequisite for deception. It was organically created to project authenticity, and thus it is another constant in the network. We suggest that it is so meaningful that it can help us map the network. It forces structure that goes beyond content delivered in what seems to be, at first glance, a shapeless and illusive activity. As such it can be complement exiting analysis tool of this phenomenon.

To do so, we utilize the machine learning technique of Random Forest Classifier (RFC)[1] to identify the patterns left by the social footprint of the fake accounts in the network [13, 14]. We exploit three databases that capture Russian influence activity on Twitter to train and test the artificial intelligence we use. The first dataset is a partly hashed[2] Twitter network that was identified and affiliated with the Internet Research Agency (IRA) (Alliance for Security Democracy data), the Russian organization that piloted the 2016 information operation [15, 16]. The second is the unhashed data on a segment of this network [4]. Finally, we use a third hashed network of Russian-speaking trolls, one that was used for domestic influence within the Russian political sphere. We train artificial intelligence on the first dataset, compare the results with elements in the second dataset for validation, and test our prediction on the third dataset.

---

[1] We preferred RFC over alternative machine learning techniques due to its specialization of reducing errors in prediction. We explored Logistical Regression, Support Vector Machine, K Nearest Nighbor, Decision Tree, and Naïve Bayes. The RFC produced superior outcomes. See Appendix Table 4, that reviews and compare methods.

The paper is structured as follows. We open with contextualizing information operations and discuss the 2016 interference case. We proceed with reviewing scholarship on the IRA operation, its network structure, and tactics. Building on those, we move to describe our theory and argument, suggesting the social qualities of the fake personas are standardized and create predictable patterns. We continue with presenting the datasets we use and our method. Next, we run our analysis on the first dataset, compare the results to the second dataset, and validate the outcome with a prediction assessment of the third dataset. We proceed with a discussion on the nature of the 2016 Russian influence network and end with a conclusion.

**Trolling instead of shooting**

The huge strides in technological development affected the way societies conduct war. Weapon systems improved with every technological leap, creating more accurate and sophisticated arms. From war machines that are controlled remotely by human operators to smart bombs with cameras that can abort a mission midflight and wait for a new target designation [17, 18, 19]. Yet, technological developments did not stop at the battlefield and expand across multiple civilian domains, creating a new social, political, and economic reality where nations, not armies, have a greater electronic impression. Personal electronic devices (*i.e.,* cellphones, tablets, smart wear, and laptops), home and car electronic features (*i.e.*, GPS systems in cars, smart home security products, virtual assistant technology, and internet refrigerators), and smart environmental technology (*i.e.*, AI integrated streetlights, AI regional water management systems, QR codes, and air pollution sensors) define the modern life experience and indicate the depth of societal dependency on integrated technology for work, consumption, travel, and recreation. This reality invited the development of cyber warfare and operations designed to disrupt and diminish the electronic infrastructure of rivals as an alternative or complementary means of war. This realm of conflict included the usage of armament designed to target communication and power arrays as well as more subtle tactics, among them jamming signals and hacking into computer systems [20, 21, 22].

---

[2] Hashed means that identifying attributes, such as username and description are hidden, replaced with a string of numbers and characters for anonymity.

Technological advances made another leap in the past few decades with the accessibility of mobile devices and social media platforms. Those innovations brought with them meaningful implications for the cyber realm. They created a social electronic infrastructure that included every person connected through personal devices. More importantly, within this new reality, every person produced a distinct digital trail, one that can be traced, studied, and manipulated [23, 24, 25]. To put it simply, in the cyber warfare realm, technological progress opened the door to "hack" people and not only computers [26]. This is where information operations and influence operations focus, assuming cyber tactics and becoming part of the cyber effort.

Information operations (IO) has been defined by the United States Government Accountability Office (GAO) as "the integration of information-related capabilities during military operations to influence, disrupt, corrupt, or usurp the decision-making of adversaries and potential adversaries while protecting our own" [27]. A RAND study commissioned by the U.S. Army underscores in its definition the holistic aspect of those types of operations and civilian dimensions [28]. The study stresses the involvement of all instruments of national power (*i.e.*, diplomacy, information, military, and economics) in the process. Generally, IO has been aimed at specific strata of experts, individuals associated with the diplomatic world, military personnel, and employees of critical infrastructure. It was part of propaganda or intelligence-targeted attempts. The 2016 Russian information operation, as well as the 2014 Russian influence operation in Ukraine, refitted that tool [29, 30]. On those occasions, it's been used to influence public opinion and the general population. Instead of targeting the diplomats and politicians, they targeted the constituents, and in the 2016 case, the process led to the election and appointment of the politicians and diplomats [ 29, 31]. This was not merely the realignment of a propaganda tool to target more people using new platforms. Instead, it underscores a profound development in warfare. Instead of influencing nations with coercive violence, adversaries can utilize cyberwarfare to degrade their rival capabilities and achieve similar outcomes as through mobilizing armies. Armies of trolls can be deployed to silence political discourse in rival countries, intervene in elections, or destabilize states [32, 33, 34]. This is an approach to the "weaponization of information" [35]. Russian theorists suggest that complementing cyber operations with hybrid tools, such as the cultivation of local paramilitary groups, can lead to "A perfectly thriving state can, in a matter of months and even days, be transformed into an arena of fierce armed conflict, become a victim of foreign intervention, and sink into a web of chaos, humanitarian catastrophe, and civil war" [36].

The 2016 Russian information operation exemplifies this approach. A Russian network synchronized an influence effort across multiple media platforms to disrupt the democratic process of the 2016 elections. It included the deployment of thousands of fake accounts across multiple social media platforms and accounts pretending to be local Americans [5, 37]. Those were animated by troll farms in

Saint Petersburg, Russia, associated with the IRA, a Kremlin-affiliated organization [38, 39, 40]. Through the wide network they foster artificial social credibility- individuals are inclined to perceive a source as credible if others do so [41, 42]. A user caught with a fabricated network of fake accounts can be manipulated to assume social credibility to dubious sources. This network promoted a campaign that attempted to influence public opinion on the democratic process, candidates, and political issues [1]. It targeted social media users *en mass* attempted to influence public discourse and promote bigotry, radical approaches, and distrust in democratic institutions, the media, and fellow citizens. It favored Donald Trump over Hillary Clinton, elevating the first while degrading the other [2, 43]. The campaign furthered controversial political agendas such as conspiracy theories, succession movements, and encouraging some groups not to vote [3, 43, 44, 45, 46, 47]. It generally fostered polarization and distrust across different crowds, stirring contested political issues such as abortion, gun laws, and police brutality while painting each topic in ever extreme colors [12, 48, 49]. Working in concert with other Russian efforts, such as "hack and dump" operation of the DNC emails on Wikileaks, the network aimed at agenda-setting for the mainstream media, fed it with provocative information and amplified the message [50, 51, 52]. The network even responded to unfolding events and tried to reorient the message, media coverage, and public opinion [53]. While the impact of this intervention on the election outcome is inconclusive, the effect it had on individuals' and groups' behavior and perceptions is well documented [15, 52].

      Trying to understand this emerging threat, multiple scholars began exploring the Russian network, its attributes, and tactics. They identify patterns in the network's behavior and structure. Several scholars explore the features of the social media platforms used, examining how the fake accounts exploited them. The number of tweets and retweets of elements is measured within the Russian network [6, 54]. They checked the number of distinct users and timed their volume of activity, every attribute of the tweeting activity has been coded and computed, including the type of URL pushed and the geolocation described by the fake users. Some researchers centered their study on one fake IRA account: @Jenn_Abrams, running a similar analysis on that specific account and looking at tweets, retweets, and replays [5]. Several studies focused on the content of the network, assessing the themes and messages delivered. Concentrating on memes depicting the presidential candidate Hillary Clinton, examined the gender framing themes and identified themes in negative visual memes [55]. Those included biological and physical character traits in addition to other gender-based stereotypes. Looking at the content and tone used by the Russian trolls, stress the language of polarization [56]. The language directed the audience toward two opposing sides of the discussion presented, pitting right and left political worldviews. The dichotomy of left versus right was further studied [57]. Their study used textual analysis to identify an emotional tone associated with left or right-leaning trolls. For example, studies discovered

that the emotions of *afraid* and *annoyance* were more frequent among right-wing tweets, while *angry* and *sad* were more widespread among left-wing tweets.

The division of left-leaning trolls and right-leaning trolls offered an additional dimension for studying the influence network, it was the first solid type of categorization of trolls' accounts. It stresses a trend in online behavior associated with specific accounts. It also indicates a role or even specialization in the network. A role that can be traced and used as an anchor in understanding the network's behavior. The right versus left trolling has been fairly common [4, 7, 39, 43, 58, 59]. This dichotomy, which can distinctly separate accounts and activity, offered researchers a structure in the shapeless and elusive troll network. Instead of a cloud of messages and fake account, scholars could clearly divide activity into two (right vs left). This stability was used as a springboard for detection. Using textual analysis tools , researchers built on this dichotomy in content to develop techniques to detect Russian trolls, separating them from regular users [8]. A different detection analysis— time-sensitive semantic edit distance that was built on the left and right content. However, instead of focusing only on the text, they focused on the actors as well [55, 58]. They identified three distinct actors, those who focus on promoting left-leaning content, those who specialize in right-leaning content, and a third group that focuses on pushing news feed content. Looking at handles in the IRA network, Linvill and Warren identified a few more actors [39]. Beyond the right and left-leaning trolls, they categorize news feeds, hashtag gamers, and fearmongers. Those categorizations offer additional springboards for researchers, allowing to further develop research and conversation around the network [3, 4, 43, 60, 61, 62].

**The Social Dimension of Fake Accounts**

The Russian network is not amorphous or shapeless. As described, scholarship identifies multiple patterns and unique attributes in the online IRA operation. Those studies identify the order in the network; an order that is dictated by an internal logic that drives the operation. Several studies offer insights into the organizational aspect of the network. Linvill and Warren present their categorization of accounts as a form of specialization in the network [4]. It means that deliberately, some accounts specialized in left-leaning trolling while others specialized in fearmongering. The specialization and hidden coordination bear similarities to the political phenomenon of astroturfing. Astroturfing is planned activity intended to create a false sense of a spontaneous and widespread grassroots movement that promotes a particular agenda while, in reality, it is originated and piloted by a hidden group or organization that uses the network as a tool. For an astroturf to work, it needs to maintain coordination across participants while simultaneously fostering false authenticity [63]. Authenticity is cardinal in the astroturf's attempt to achieve its goal. The audience would not be persuaded by activists who support maintaining the local polluting factory if they do not look native or if they are carrying the exact message the factory owners push to the media. Starting as a grassroots tactic that involves protestors and activities, astroturfing developed, adopting new tools and tactics. Protests and campaigns can run on the Internet as well. The spread of social media activism opened the door for online astroturfing [64, 65]. Studying political astroturfing on Twitter focused on the coordination of a disinformation campaign. Their study examined the South Korean National Information Service's (NIS) disinformation campaign during the 2012 presidential election, illustrating the coordinated effort of disseminating information and framing messages. Contextualize their study in the 2016 Russian information campaign, illuding for a similar rationale of coordination [66]. The Russian network's coordination was evident in its relations with Wikileaks. Wikileaks, a non-profit organization that specializes in publishing classified information provided by anonymous sources, has been working, knowingly or unknowingly, with the Russian campaign. During the 2016 Election season, Wikileaks regularly released classified race-related information. Information shared by Russian intelligence agencies [1]. The Russian troll network coordinated its operations in sync with Wikileaks releases, setting the media's agenda and framing the stories [52].

Beyond coordination, the Russian network demonstrated an active attempt to maintain authenticity. American audiences will find a Saint Petersburg resident that discusses urban grievances of the local black community in Atlanta as irrelevant or odd. It means that the trolls had to have a disguise, pretending to be fellow Americans, simulating the image and vernacular of young black activists, opinionated pro-guns housewives, conservative veterans, and others. Examination of trolls' accounts underscores the adherence of Russian operatives to the boundaries of their fake personae. A persona of eco-enthusiast would not promote pro-guns positions. Instead, it would be used to push support for the Green Party candidate, Jill Stein. The fake person's @*samirgooden* support of the Green Party "AND WITH 5% OF THE VOTE THE GREEN PARTY WILL BE ON THE BALLOT AS A RECOGNIZED PARTY IN 2020-BY PETITION IN 48 STATES NOW-THE STORM HERE-US https://t.co/EX98mBxLrj" is consistent with its critique of Texas Senator Cruz on climate denialism "@SenTedCruz's climate denial hearing is dangerous & irresponsible. We must take bold action to stop climate change. https://t.co/2BCbiDKgML." Another aspect of those accounts' effort to maintain authenticity was identified in what Linvill and Warren called "camouflage" [4]. Camouflage in the Russian influence network refers to the "noise" that accounts manufacture—messages that do not promote the Russian agenda. Instead, those messages foster credibility and sell authenticity. They are an attempt to convince the audience that the specific account they view does not have a hidden agenda and that it is legitimate, representing real people and organizations. Camouflage in the network looks like messages saying "#ThingsYouCantIgnore Alcoholism," "Mondays #MustBeBanned," and "The Tuna-night Show #FishTV @midnight" written by the user @*traceyhappymom*. Those messages do not carry an agenda or dive into controversial issues. Instead, they aimed to solicit sympathy by talking about a relatable issue such as alcoholism, hating Mondays (a common American reference), and discussing TV shows. The general theme behind camouflage messages is the attempt to portray a well-rounded character, one that is not exclusively focused on politics or wedge issues. The troll network invested much time and effort in camouflage. Camouflage messages represent a meaningful portion of the trolls' activity, with Linvill et al. marking it as over 50% [4].

Building on the specialization, authenticity, and camouflage functions, we assert that the social dimensions of the fake persona constrained the trolls' behavior, forcing it to follow specific lines. While Linvill and Warren focused on functions in the networks, here we focus on the presentation of the accounts. An account that pretends to represent a local news outlet from a town in rural Missouri cannot share tweets on its mood this morning or how a girl broke his heart. This is true for other online behavior, such as hashtags used or whom you retweet. A news outlet would not retweet an anonymous person or use an offensive hashtag. Doing so will break its authenticity. Accounts are not fluid in their representation. An account cannot alter its portrayal overnight, shifting from being an organization to an individual without losing credibility and risk detection. A "slip up" of an account invites detection, something trolls are trying to avoid. Those confinements force a social footprint that is stable. This stability allows us categorization of the accounts by type and can even be used as a springboard for the prediction of the trolls' behavior. We suggest that an account that pretends to be a person, an individual, cannot pretend to be a news outlet the day after. The social footprint of an individual is different from that of a news organization. Linvill demonstrates that the trolls used different types of camouflage for different actors [4]. Camouflage for an organization is different from the one for a person. Breaking the camouflage unmasks the deception and harms authenticity. Thus, we theorize that, for the most part, the trolls will adhere to their assigned camouflage per their social footprint. Furthermore, we assert that the distinct categorization will manifest in more than content. It will be treated differently in the influence network and will have a unique pattern of online behavior (for instance, volume of tweets or number of replies). For instance, we can expect that Organizations or News Outlets will reply less than Political Affiliate accounts.

We categorize four types of social footprint among the Russian trolls in order to differentiate troll activities based on the latest research in the field, especially that of Linvill and Warren [4]. The proposed distinct conceptual categories are logically non-overlapping groups, so each account can be placed into one category without ambiguity. Further, the selected categories are relevant to the research question and have a statistically significant impact on the outcomes being studied while machine learning algorithms can reliably distinguish between accounts belonging to proposed criterion. The first two categories are those that pretend to be organizations and the latter two pretend to be individuals. The first categorization is the *Fake News*, referring to the Russian operatives masquerading as local news outlets (*i.e.,* "Kansas City Daily News," "Memphis Online," "Pittsburgh Today"). These account descriptions foster the image of a real news outlet instead of a news enthusiast or an individual that posts news on a topic (see Appendix, Table 5). This is where the Fake News category differs from the News Feeds function that was identified by other scholars [39, 58, 59]. For example, the account @TEN_GOP, which is considered one of the most prolific News Feeds in the network, describes itself as the "Unofficial Twitter of Tennessee Republicans. Covering breaking news, national politics, foreign policy, and more." It does not pretend to be a new channel or a newspaper but rather a group of people (or a person) associated with a political party that covers news. This description is substantially different from the Kansas City Daily News of: "Local news, sports, business, politics, entertainment, travel and opinion for Kansas. DM us 24/7." In this case, the trolls attempted to present a local news outlet. Nothing in the description illudes that it is run by individuals or news junkies. Thus, this account generates specific expectations that are associated with the online behavior of established news outlets. They are expected to present a more neutral voice in their messaging and bring up a broad array of mundane news on the weather, traffic, sports, crime, and others. They cannot get into an argument with other accounts and cannot share their feelings or describe how their morning was. Masquerading as news outlets demand a specific tone and type of online activity—a unique social footprint.

The second category is *Organizations*. This type of account imitates social organizations and not individuals. Here we can find social movements, nonprofits, and companies. Categorically, organizations behave differently from individuals online. They do not share personal stories, they do not deviate from their specific agenda, and they are invested in the organization's goal. Organizations are social entities created for a specific purpose. Some of them can be very sophisticated and involve multiple stakeholders and complex bureaucracy. Others are more basic and can include a limited number of individuals. Nonetheless, organizations existence is contingent on their goal—the reason they were formed. This reason, whether it is to sell a product, support a specific community, or create art, dictates the organization's behavior, online or otherwise. Examples of organizations in the network are the *@MatEvidence*, described as the "Official Twitter account for Material Evidence photo exhibition," or *@BlackToLive*, which put in their description, "We want equality and justice! And we need you to help us. Join our team and write your own articles! DM us or send an email: BlackToLive@gmail.com."

The third and fourth categories focus on individuals. We identify two types of social footprints for individuals. Individuals in the Russian network present themselves either as *Political Affiliates* or as *Default Individuals*. The Political Affiliates accounts are those that overtly focus on political issues and debates. They are political animals and the crusaders for a political side of a debate, whether it is the Black community or pro-Trump conservatives. Their self-representation is very clear, allowing the audience to understand they are getting into a political debate. An account that description states, "Freedom is never given; it is won. #BlackLivesMatter", "Calm down, I'm not pro-Trump. I am pro-common sense. Any offers/ideas/questions? DM or email me jennnabrams@gmail.com (Yes, there are 3 Ns)", or simply "#NeverHillary." These accounts' political tone is the reason for followership. We can expect a politically overt account to extensively discuss politics. Partisan political debate, commentary, and humor are what those types of accounts are expected to produce. Though they can post personal messages, these accounts are expected that a significant portion of their content will focus on politics and on a specific side of the debate, they cannot reliably switch sides. This is their social footprint. This is in contrast to Default Individuals, the fourth category. It is the default category. Those accounts do not present any political affiliation in their description. They exhibit accounts that are indistinguishable from real users, by description at least. Their descriptions are random, discussing sports, personal characteristics, or nonsense and silly verbiage. Yet, all across, they present an individual, not an organization, and not a politically focused individual. This is their footprint.

We argue that those four categories of social footprint are meaningful enough to force a structure on the fake accounts. Notice that our entire sample is comprised of identified fake accounts, we do not attempt to present a detection tool in this study. Instead, we assert that this structure of categorization offers a springboard for predictive analysis of network behavior on pre-identified fake accounts behavior or as a tool that can be developed into detection.

**Method and Data**

To illustrate the persistence of the Russian networks' social footprints, and therefore its usefulness for future analyses, we generate a prediction of the categorization and test its validity on two different datasets. We use machine learning to first identify how each category's social footprint manifests in activity and second, based on that, to produce the prediction. Several scholars used artificial intelligence tools to analyze the Russian network. Behavioral and linguistic parameters have been used with a variety of machine learning tools to differentiate accounts for trolls and non-trolls [9]. Another research proposed using Inverse Reinforcement Learning (IRL)to identify troll accounts and troll behaviors, arguing that this model can classify troll accounts from non-troll accounts using the accounts identified by the US Congress regarding the 2016 U.S. presidential election [67]. An emotional and lexical-based technique have been proposed to identify trolls that operated during the 2016 U.S. presidential election [8]. The researchers considered a few prominent features, such as account topic and other profiling features to discover the trolls' real identities. Another content-based analysis has been introduced to identify influence operations on social media using a set of classifiers [69]. Others used a machine learning approach to find the behavioral patterns of trolls according to their roles, such as left troll, news feed, and right troll [68]. They used traditional supervised learning and distant supervision frameworks for labeled and unlabeled trolls. They validated their methodologies on the IRA dataset.

Continuing this body of research, we train our artificial intelligence on a sample, linking the categorization with their unique activity. We take a five-stage analysis. First, we manually coded categorization in a sub-sample (about 65%) of Twitter accounts that are associated with the IRA and written in English. The categorization is based on the accounts' description. When the description was missing (*i.e.*, hashed), we used instead repeating hashtags in the content, randomly sampled, that are associated with each type of categorization (*i.e.,* #MAGA for *Political Affiliates*, #NEWs for *Fake News*). By doing so, we included another 21% of the sample and increased the training sample for the AI.[3] Next, we train the machine learning model to identify those categories based on their unique activity. Here we look at a set of features that includes the frequency and timing of tweets and retweets, along with other account and activity features (see Table 2). In the third stage, we use machine learning to predict the categories in the rest of the sample, the hashed part where we cannot assess the account's category, and that does not have hashtags in its content. The fourth is the validation stage, where we check if our prediction of categorization of the hashed data is valid. Here we compare our categorization results to Linvill and Warren's coding, looking for similarities and gaps [39]. Lastly, we conduct a second and more comprehensive validity test on a sample of Russian tweets written in the Russian language. Here we started with the prediction and checked the results manually using the users' description.

Our analysis used three Twitter datasets of IRA activity, one for training the artificial intelligence and two others for validation (1. IRA English, 2. IRA Russian, and 3. Linvill and Warren) [4]. The main dataset, published by Alliance for Security Democracy data (https://www.io-archive.org/#/),consists of about ten million tweets in 58 languages (Table 1). We can recognize forty features in the dataset, among them profile-related features, qualitative behavioral measures, and tweet-related linguistic features. For privacy reasons, part of the dataset is hashed, meaning users do not have user id, user display name, user screen name, and user profile description. Instead, they have a hashed number for all features. We extract the English and Russian tweets, constructing two separate datasets: The IRA English (2,992,134 tweets) and the IRA Russian (4,853,185 tweets). The IRA English dataset has 2,832 unique users and covers the period between November 2009 to May 2018. About 64% of those accounts are unhashed, namely, we have readable profile descriptions. The IRA Russian dataset consists of 1,554 unique users and covers the same timeframe. The Linvill and Warren dataset (1,875,029 tweets) used cover IRA accounts' activity between June 2015 to December 2017, as identified by the 2018 release of the U.S. House Intelligence Committee.

---

[3] The predictive results of training the AI on 64% of the sample did not yield high accuracy. As such, we used this method to inflate the training sample.

**Table 1** Datasets Features

| Dataset | # Users | # Tweets | # Hashed Users | Timeframe | Languages |
|---|---|---|---|---|---|
| Alliance for Democracy Dataset | 5,074 | 8,768,633 | 1,232 | November 2009 to May 2018 | 58 Languages |
| IRA English | 2,832 | 2,992,134 | 1,019 | November 2009 to May 2018 | English |
| IRA Russian | 1,554 | 4,853,185 | 119 | November 2009 to May 2018 | Russian |
| Linvill and Warren | 1,133 | 1,875,029 | 0 | June 2015 to December 2017 | English |

To generate the categorization, we examine the accounts' names and descriptions. This could be done only on the unhashed accounts. We conducted two coding rounds on all the unhashed accounts in the sample, achieving intercoder validity of over 95%. Accounts portrayed as news outlets were marked as *Fake News*. Accounts that were depicted as organizations or social movements were marked as *Organizations*. Accounts described as individuals invested in political discourses (right or left) were identified as *Political Affiliates*. The rest of the accounts, which portrayed individuals, were categorized as *Default Individuals*. Table 2 captures the descriptive statistics of the four categories. It shows that *Organizations* and *Fake News* represent a smaller share of the network's activity (101 and 136 respectively).

**Table 2** Descriptive Statistics

| Features | Fake News | Organizations | Political Affiliates | Default Individuals |
|---|---|---|---|---|
| N | 136 | 101 | 595 | 2,000 |
| # Tweets | 91,539 | 158,605 | 323,714 | 1,337,698 |
| # Retweets | 78,036 | 44,529 | 162,419 | 795,594 |
| # Users Mentioned | 119,093 | 63,172 | 594,701 | 1,425,592 |
| Avg. Followers | 9,137.65 | 27,593.52 | 5,126.85 | 11,339.62 |
| Avg. Following | 3,858.04 | 6,364.18 | 3,144.11 | 5,601.21 |
| # Hashtags | 116,919 | 129,838 | 273,813 | 1,474,905 |
| # Replies | 5,348 | 7,508 | 196,641 | 1,490,090 |
| # Likes | 39,073 | 78,892 | 6,487,172 | 23,487,246 |

Building on this categorization, we examine the unique features distribution of each user, identifying similarities per category, such as the volume of replies, the frequency of tweeting, and other features (Table 2). We suggest that each category presents a unique social footprint that, in turn, manifests in specific patterns of activities. Namely, the rhythm and volume of tweets and retweets of political affiliates as a group will be different from that of organizations.

At that point, we have the unhashed data that includes a description and that we coded as one of the four categories (n=1813). We also have the hashed data that we could not code (n=1019), given that they do not have a description. To further increase our sample to allow the effective training of the AI, we use their hashtag footprint (the hashtags that are associated with the categories we manually identified) to further decipher their categorization, Table 6 in appendix.

Often hashtags are time-sensitive, communicating with specific events and spike during a limited time frame. Thus, if we consider year-by-year subspans, the frequency distribution of those hashtags may not represent the actual hashtag usage, and it provides an underestimation of the hashtag utilization. At the same time, considering monthly subspans will lead to over-estimate hashtag utilization. Therefore, on average, we divided the dataset into six months subspans, $sp = \{sp_1, sp_2, \ldots, sp_n\}$, based on their tweet time, and then filtered all hashtags for each user within the corresponding subspan. To proceed with hashtags, we use Natural Language Toolkit (NLTK) 50 corpora and lexical supports in Python to process the features [70]. Given the Russian network is dynamic and changing, we use the library created by the NLTK on six months subsets, with each analyzed separately [71]. The NLTK helps us identify a collection of vectors that are associated with each user.

Next, we created matrices using hashtag usage for already categorized user accounts and uncategorized accounts, which have no readable profile description, to identify these uncategorized accounts in each subspan. We use the vectors between users and their hashtag usage on the unhashed sample as a blueprint to identify categories. For example, we can see that user "Rita_Hart" used the hashtags "IHaveARightToKnow," "ItsRiskyTo," and "GiftIdeasForPoliticians" but did not use the hashtags "sports," "entertainment," "news," and "health" within a given subset. We compare this blueprint to the hashed account. We match patterns of hashtag usage and assign the fitted categories. A matrix of the uncategorized users by $U$, as in:

$$U_{mn} = (u_{.1} \quad \ldots \quad u_{.n}),$$

where $m$, $n$ represent the total number of unique hashtags and the number of uncategorized actors in the current time span, respectively. Similarly, the matrix of categorized actors denoted by $V$, as in:

$$V_{mk} = (v_{.1} \quad \ldots \quad v_{.k}),$$

where $k$ represents the total number of categorized actors in the current time span. Next, we can calculate the similarity between $u_i$'s and $v_j$'s in a subspan via,

$$\text{similarity}\,(u_{.i}, v_{.j}) = \frac{u_{.i} \cdot v_{.j}}{||u_{.i}||\,||v_{.j}||},$$

where $i \leq n$, and $j \leq k$ [72, 73]. Next, we assign an impermanent category to $u_{.i}$ of the corresponding actor $v_{.j}$ by considering the maximum similarity $(u_{(.i)}, v_{(.j)})$. We continue this process until the categories for all the $u_{.i}$'s (hashed users) are approximated in the current subset. This process is recursive until we determine the categories of all uncategorized users in all the subsets. However, one particular uncategorized actor may appear in more than one of the subsets, and its impermanent category may change from one subset to the next.

To solve the impermanent category problem, we list the relative frequency distributions for all uncategorized actors regarding their impermanent categories across the subsets and the highest relative frequent category assign to the particular actor. Eventually, no more than one mode (the highest frequent category) appears in each relative frequency distribution during the calculation. Through this method, we approximate categories for all of the uncategorized users in this dataset (see Fig. 1). The accounts with workable hashtag footprint account for another 21% of the sample (n=595). This means that through this process, we increase our training sample from n=1,813 to n=2,408 while reducing the uncategorized group from n=1,019 to n=424.

**Fig. 1** The process of identifying hashed actor categories (595 accounts) in the 1st dataset according to the four conceptual categories

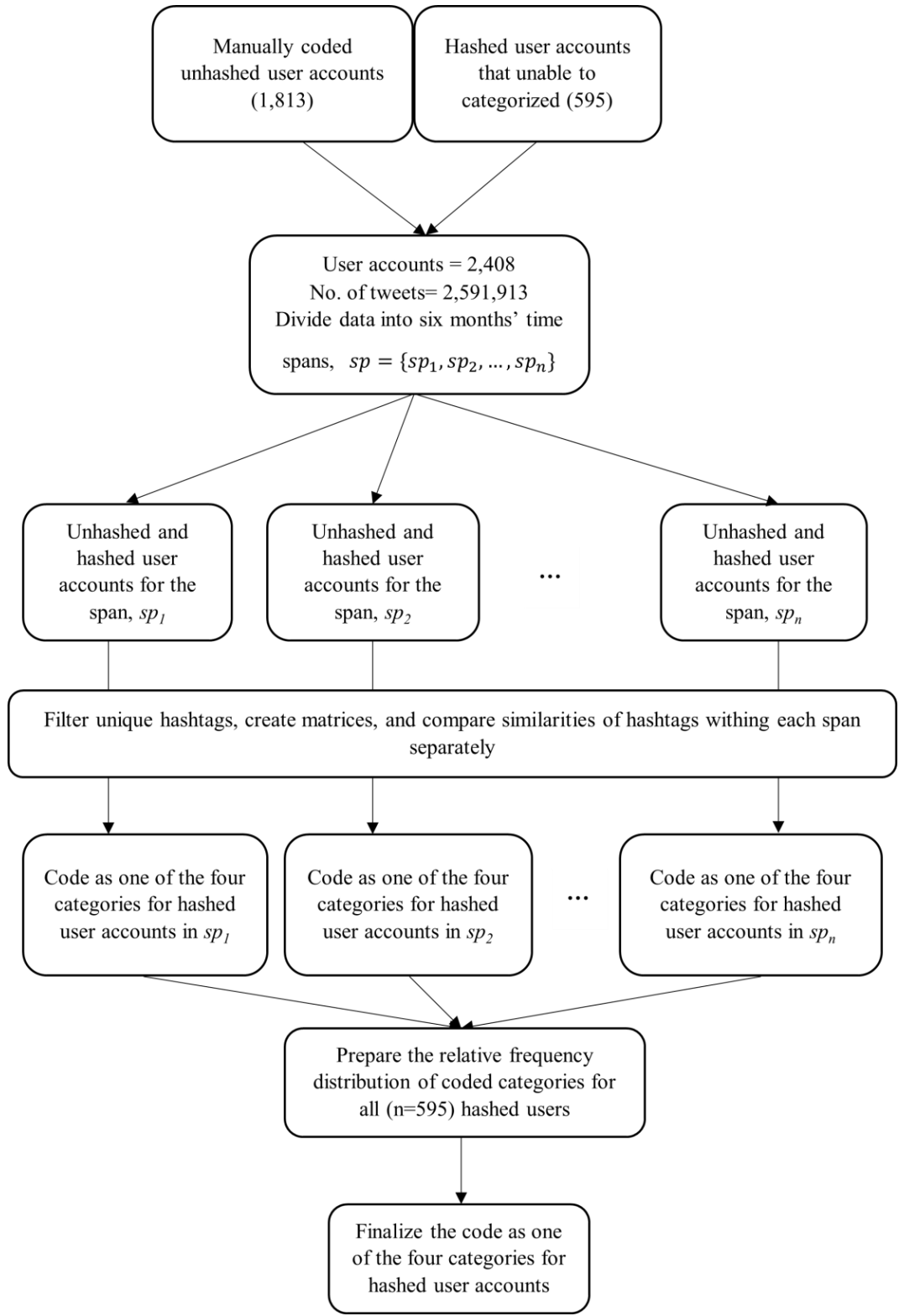

**Predictive Model**

After identifying 85% of the samples (n=2,408) in the English tweets network and assigning them to one of the four conceptual categories, we move to the predictive stage, to test the stability of the categorizations. We use RFC for the prediction, a machine learning algorithm that uses multiple decision trees, which help reduce errors in prediction [13, 74]. To overcome multicollinearity, we identify the eight features that minimize multicollinearity based on Pearson's correlation coefficient test [75]. We feed the RFC with those eight features (see Table 2) that are associated with users and categories, teaching it to link the unique behavior per feature (the social footprint) with particular categories. Adding the categorization created an imbalance in the dataset, an imbalance that the RFC is designed to compensate for. To preserve the structure of the dataset, we performed the Least Absolute Deviation (L1) normalization making sure the weights of each variable are aligned [76, 77]. To train the artificial intelligence, we divide the dataset (n=2,408) into two subsets: training (70% of the sample) and testing (30% of the sample).

The RFC is a collection of many decision trees that keeps the minimum relationship among trees and by doing so, reduces prediction errors (see Fig. 2 for training process). The training data (1,685 user accounts) divide into $n(=100)$ subsets randomly. Each subset consists of a training set, which is less than 1,685 samples and 723 testing samples, and trains an algorithm using a decision tree to predict categories. Each tree trains its' own algorithm independently based on its' subset. The algorithm is based on all decisions from every tree. Fig. 3 illustrates the performance change with the depth of a randomly selected tree. The RFC aimed to achieve the optimum depth of a tree given that an overelaborate tree will overfit and too few levels in the tree will underfit the model. Fig. 3 illustrates these dynamics for a tree, with the best fit achieved at seven levels of the tree (86.48%), while the lower and higher levels produce less accurate predictions.

**Fig. 2** The flow chart of the process of building an RFC algorithm

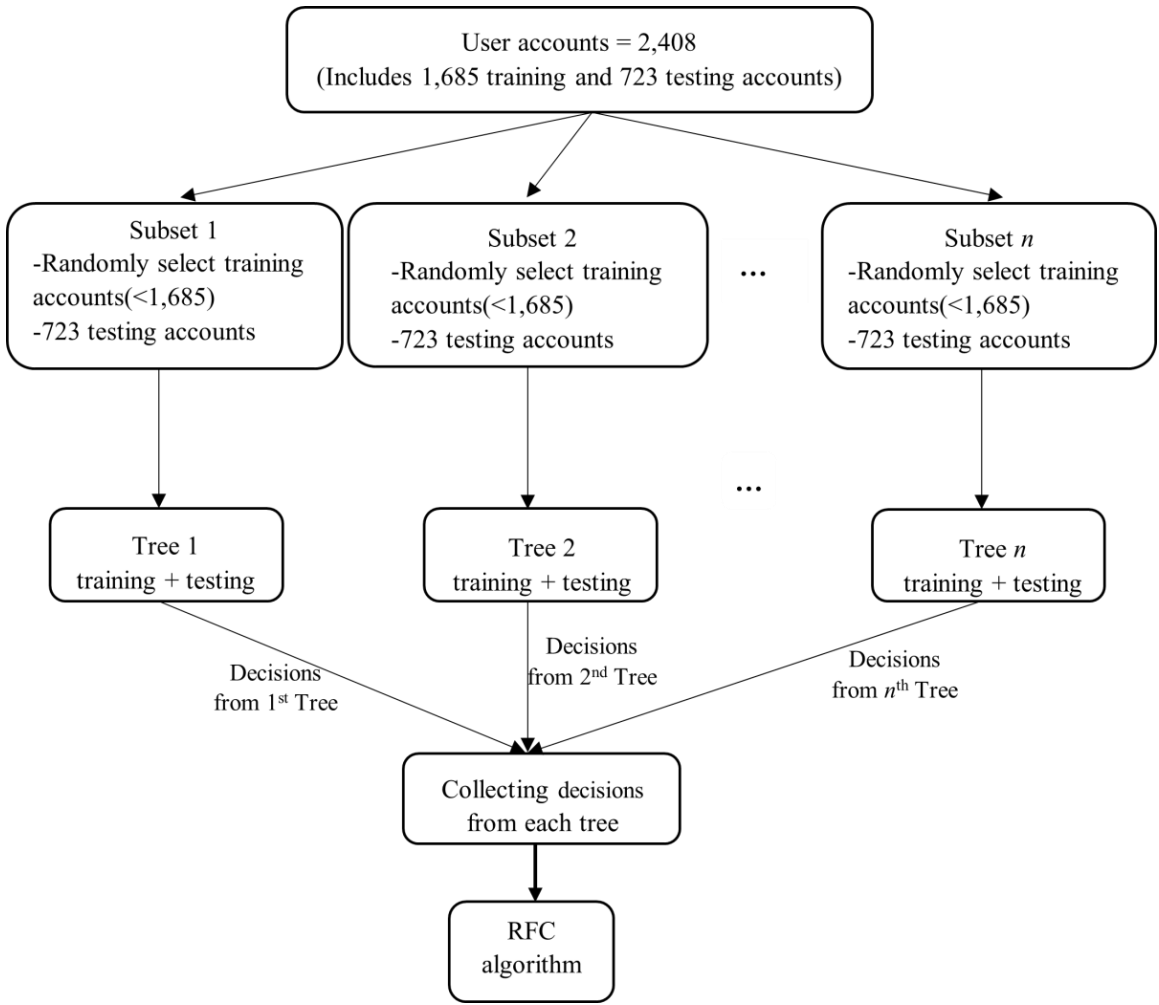

**Fig. 3** Accuracy changes with the depth of a randomly selected tree

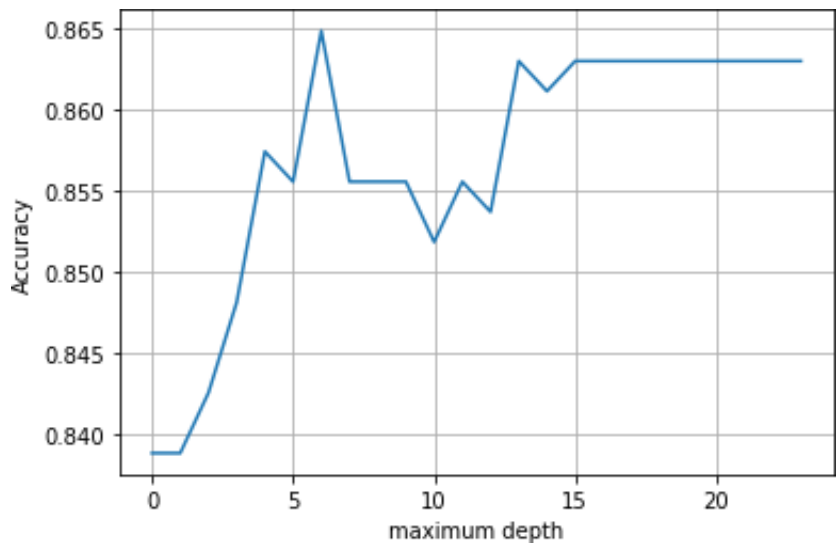

The RFC uses the Gini index process multiple times on the training dataset until it achieves optimal classification [78, 79]. It controls for the greatest possible depth of the decision tree by expanding nodes until all leaves become pure. At that point, the model is optimized, using the best classification to produce a predictive output. For prediction validation, the RFC runs a stratified fivefold cross-validation on the output from the test set. This process offers the model's predictability accuracy. We use a fivefold cross-validation and higher due to the sample's size of the Organization category. Greater fold cross-validation would delude the Organization category and manufacture a skewed prediction. Running our model on the training dataset, we achieve an output that stands at 88% accuracy; this is our prediction.

To evaluate our predictive model's accuracy per category, we use three measurements based on sensitivity and specificity test that we run on the test set. The measurements are precision, recall, and F1-Score [80, 81]. The results based on those three measurements capture the model accuracy, not only the overall accuracy but the individual categories as well. Table 3 shows that all categories enjoy relatively high levels of accuracy across the three measures. The categories of Fake News and Default Individuals present the highest levels, while Organizations show the lowest.[4]

**Table 3** Predictability Measures for Each Category Based on Precision, Recall, and F1-Score on the Test Set

| Category | Precision | Recall | F1-score |
| --- | --- | --- | --- |
| Fake News | 0.94 | 0.82 | 0.88 |
| Organizations | 0.82 | 0.80 | 0.81 |
| Political Affiliates | 0.83 | 0.85 | 0.84 |
| Default individuals | 0.88 | 0.96 | 0.92 |

**Predictive Model Validation**

To test the predictive model, we go beyond our dataset, using two datasets for validation: the Linvill and Warren dataset and the IRA Russian language dataset. Linvill and Warren offered a systematic classification of IRA activity around the 2016 Russian interference operation. Working with a dataset of about two million tweets associated with IRA accounts they identify five types of categories: right-leaning trolls, left-leaning trolls, news feeds, hashtag gamers, and fearmongers. Their dataset was shared online, including the users' categorization coding. Their categorization is based on the interplay between tweets' content and profile description and name. Our validation test focused on one of those categories, the News Feeds. Of this dataset's categorization, News Feeds represent the closest fit to our categorization of Fake News. The two are not identical, with Fake News focusing only on accounts that impersonate news outlets, while Linvill and Warren also include individuals that specialize in pushing news.

---

[4] Organizations show the lowest due to their relatively small sample size.

We use the Linvill and Warren classification of News Feeds to check the validity of our prediction due to its high similarity to our category of Fake News. We compare the accounts identified by Linvill and Warren as News Feeds to those our RFC identifies as Fake News. We do not expect identical results. Yet, we do expect to see a high correlation indicating that Linvill and Warren's classification is close enough to our artificial intelligence's model prediction. We start the assessment with comparing the two datasets and categories to similarity in users. The IRA English is larger, with 2,848 users versus 1,133 for the Linvill and Warren dataset. It also covers a longer timeframe, November 2009 to May 2018, versus June 2015 to December 2017 (see Table 1). Thus, we expect that the IRA English database will have more users and activity than its counterpart. The News Feeds category includes 54 unique users for Linvill and Warren dataset. In contrast, the Fake News category in the IRA English dataset includes 136 unique users. Examination of matching accounts across the two datasets shows that there are 49 users in those two groups that are identical; we focus on those for the validation process. Within the identical samples, we observe that forty-nine Fake News accounts match with the News Feeds users, representing 90.7% accuracy for the model prediction. To put it simply, the RFC model that focused on fake news profiles was able to reach a match of about 90% when tested against the Linvill and Warren's manual coding.

For stronger validation, we test our model on a parallel dataset—the IRA Russian-speaking dataset (the cleaned dataset can be found in GitHub: https://github.com/SachithDassanayaka/IraTweet). We run our model on the dataset and then manually code the categories based on description and username, reversing the process. Aiming for a domestic audience and in collaboration with the Russian government, this dataset offered a new type of social behavior. On some occasions, some users in the network masqueraded as government officials and functions, for example, the Chairman of the Youth Council in a local district (Председатель Молодежного Совета района Чертаново Северное. Член ЕР). This social behavior is not possible without government consent. Given that in the Russian context, those types of accounts do not represent a person but an organization, a government-affiliated organization, we marked those as Organizations in our manual coding. To validate the accuracy of the predictive model, we compare similarities between the results of the RFC predictions and the manual coding categories. The validation results through the Russian language tweets subset indicate that our predictive model achieves overall 90.5% prediction accuracy.

To summarize, training the model on the IRA English data offer 88% accuracy in prediction categories across accounts. Comparing one category of our prediction, the Fake News, against the Linvill and Warren dataset fitted category, News Feeds, demonstrates a 90.7% accuracy. Testing our social footprint prediction on the IRA Russian exhibited 90.5% accuracy. Essentially, those results indicate that the trolls accounts have a social footprint, one that is forced by social expectations and that produces a coherent and distinct set of indicators that can be used for prediction, analysis, and in the future for detection.

**Conclusion**

Does Russian troll have a social footprint? This study suggests that they do. The IRA trolls' pursuit of authenticity to convince their audiences forced them to adopt social norms and expectations. They were required to follow social expectations and present themselves as legitimate social actors. By doing so, the troll factory developed patterns of behaviors associated with different types of social categories that are identifiable. Our analysis shows that the social footprint is solid enough to allow us to predict users effectively by social categories. We trained artificial intelligence on the operationalization of these social footprints, assigning users to four distinct categories. We tested the validity of our prediction on two datasets of IRA networks on Twitter. Our predictive model achieved high validity, testifying to the robustness of the social footprint.

The contribution of this manuscript goes beyond a sociological exercise, demonstrating how social forces dictate the way we manipulate one another. This study's main contribution is in adding a mark for information operations. Scholarship that focused on patterns within the network helped us better understand those operations. It facilitates mapping actors and their activities. It also offers us methods and ways to detect their activity. We add the social footprint to the list of other footprints used to identify and predict trolls' behavior, among them left and right political affiliation, categorization by function, and textual analysis of content. The robustness of the social footprint, as tested and illustrated in this study, presents an additional measure to an elusive network. Following those findings, future study should venture to content analysis in line with this categorization or for detection.

Information warfare, and within this influence operations, turned to be the weapon of choice in the great power competition [82]. It is a low-cost and high-yield tactic that is not regulated well by international law, and thus hard to deter [31]. Effectively, it is challenging to identify the attacker, and even if known, the retaliation is not certain or tolerable. The IRA interference operation in the 2016 American Presidential Election is not a one-time event. Russian operatives, including the IRA, continued and continuing their effort to shape public opinion in the U.S. and other countries [83, 84]. Russia is not the only actor utilizing the developments in the cyber realm. China has been running several information operations, trying to influence global public opinion over the Hong Kong demonstration or incite physical protests around COVID-19-related concerns in the U.S. [85, 86]. Iran targeted anti-Trump voters, pretending to be members of far-right American groups, like the Proud Boys, and sending online threats with the hope of causing outrage and mobilizing these voters [87]. Sparring across the cyber realm is here to stay. A better understanding of those operations and developing new ways to identify, map, and detect those activities is a priority.

**Conflict of Interest**

The authors confirm that there are no known conflicts of interest associated with this publication and there has been no significant financial support for this work that could have influenced its outcome.

**Author contributions**

All authors contributed to the study's conception and design. Material preparation, data collection, and analysis were performed by Dr. Ori Swed, Dr. Dimitri Volchenkov, and Dr. Sachith Dassanayaka. The first draft of the manuscript was written by Dr. Ori Swed, and all authors commented on previous versions of the manuscript. All authors read and approved the final manuscript.

**Data availability**

This research used a publicly available dataset which can be downloaded from the Alliance for Security Democracy data website. https://securingdemocracy.gmfus.org. Further, we have uploaded some of the datasets that we created during the training and validation, and you can find them through, from GitHub: https://github.com/SachithDassanayaka/IraTweet

**Disclosures and declarations**

There are no sources of funding or any financial support for this study.

**Appendix**

**Table 4** Average accuracy for selected classifiers

| Classifier | Average Score |
| --- | --- |
| Logistic Regression | 0.26 |
| Support Vector machine | 0.23 |
| K Nearest Neighbor | 0.69 |
| RFC | 0.88 |
| Decision Tree | 0.44 |
| Naïve Bayes | 0.45 |

**Table 5** Manually coded actor categories based on their user profile description

| User profile description | Category |
|---|---|
| Official Twitter account for the online magazine "United Muslims of America," a page to unite all Muslim people living in the USA. | Organizations |
| Newarks latest news source. Follow us for original reporting and trusted news @NewarkVoice | Fake News |
| Political & Military Analyst | Political Affiliates |
| Every child is an artist. The problem is how to remain an artist once he grows up. | Default Individuals |
| Social media enthusiast. Wannabe creator. General internet expert. Entrepreneur. Hardcore introvert. | Default Individuals |
| Non-Governmental Organization (NGO) | Organizations |
| Breaking news, weather, traffic and more for New Orleans and Louisiana. DM us anytime. RTs not endorsements | Fake News |
| warm as the sun, dipped in black. full time protagonist, womanist, & revolutionary | STRENGTH, COURAGE, & WISDOM. | Default Individuals |
| Welcome to the official Department of Space Twitter account! | Organizations |
| Constitutional conservative-Pro life-Pro 2nd amendment-Christian | Political Affiliates |
| When in danger or in doubt, run in circles, scream and shout | Default Individuals |
| Local news, sports, business, politics, entertainment, travel and opinion for Detroit. DM us 24/7 | Fake News |
| We are a club of people who love New York City. Follow us or visit our website, then you can know everything about NYC. | Organizations |
| Houston top news and stories, powered 24/7 | Fake News |
| Conservative; Right and proud; Christian. Love my country and will stand against liberals and socialists. | Political Affiliates |

**Table 6** Hashtag based actor categories for actors who have hashed user profile description

| Hashtags | Category |
|---|---|
| #CityOfChester #Chester #Pennsylvania #USA #America #US | Organizations |
| # TopNews # News | Fake News |
| #Moscow | Political Affiliates |
| #Music | Default Individuals |
| #ToDoListBeforeChristmas | Default Individuals |
| #DefundObamacare #StandWithCruz #MakeDCListen #repealobamacare #TCOT | Political Affiliates |
| #Secede #Texas #Secession #Austin #Texan | Organizations |
| #RostovNaDon #Rostov #News | Fake News |
| #MutualFollowing | Default Individuals |
| #BlackLivesMatter #ACAB | Organizations |
| #Christian, #American, #Patriot, #TeaParty, #TGDN, #TCOT, #NRA, #LNYHBT, #PJNET | Political Affiliates |
| #WakeUpAmerica #tcot #RedNationRising | Default Individuals |